\documentclass[aps,prl,12pt]{revtex4-1}

\usepackage{graphicx,graphics,color,epsfig}
\usepackage{amsmath}
\usepackage{amssymb}
\usepackage{amsfonts}
\usepackage{amsfonts}
\usepackage{epstopdf}
\usepackage{bm}
\usepackage{times,xspace}
\usepackage{color}

\newcommand{\blue}{\textcolor{blue}}

\begin{document}

\title{Engineering three-dimensional topological insulators in Rashba-type spin-orbit coupled heterostructures}

\author{Tanmoy Das$^{1*}$, A. V. Balatsky$^{1,2,3}$ \\
\normalsize{$^1$Theoretical Division, Los Alamos National Laboratory, Los Alamos, NM 87545, USA}\\
\normalsize{$^2$Center for Nanotechnologies, Los Alamos National Laboratory, Los Alamos, NM 87545, USA.}\\
\normalsize{$^3$Nordita,  Roslagstullsbacken 23 , 106 91 Stockholm Sweden.}\\
\email{Email: tnmydas@gmail.com.}}
\date{\today}

\begin{abstract}
{\bf  Topological insulators represent a new class of quantum phase defined by invariant symmetries and spin-orbit coupling that guarantees metallic Dirac excitations at its surface. The discoveries of these states have sparked the hope of realizing nontrivial excitations and novel effects such as a magnetoelectric effect and topological Majorana excitations. Here we develop a theoretical formalism to show that a three dimensional topological insulator can be designed artificially via stacking bilayers of two-dimensional Fermi gases with opposite Rashba-type spin-orbit coupling on adjacent layers, and with inter-layer quantum tunneling. We demonstrate that in the stack of bilayers grown along a (001)-direction, a nontrivial topological phase transition occurs above a critical number of Rashba-bilayers. In the topological phase we find the formation of a single spin-polarized Dirac cone at the $\Gamma$-point. This approach offers an accessible way to design artificial topological insulators in a set up that takes full advantage of the atomic layer deposition approach. This design principle is tunable and also allows us to bypass limitations imposed by bulk crystal geometry.}
\end{abstract}

\maketitle

The unusual properties of topological insulator (TI) rely on synthesizing suitable materials that inherit various invariant symmetries and spin-orbit coupling in the bulk ground state, and that allow the formation of metallic Dirac fermions at the boundary.\cite{Hasanreview,Zhangreview,FuTCI,RoyZ2,Schnyder,Zaanen} From the theoretical standpoint, the bulk symmetries include a combination of low-lying odd parity orbitals, and time-reversal symmetry with a crystal geometry that can conspire an inverted band structure via strong spin-orbit coupling,\cite{FuKane07,Bi2Se3SCZhang} or crystal mirror symmetry.\cite{FuTCI} The main focus of the search for TIs so far has been limited to the synthesis of bulk materials with these inherent characteristics. Compounds of choice are those with a large bulk spin-orbit insulating gap (Bi-based compounds\cite{Hasanexp,ZXShen} and their functional variants\cite{HueslerHsin,HueslerZhang,Li2AgSb,FuTCI}). The efforts have then been extended to manipulating the `non-trivial' topological phase by driving a trivial topological system through a topological phase transition via chemical doping,\cite{TIPTHasan,Bi2Se3doped} by tuning the lattice constant,\cite{Li2AgSb,HgTestrained} or by introducing broken symmetry quantum phases.\cite{TMI,TAFM} These materials offer tremendous opportunities for applications as well as realizations of numerous nontrivial properties such as  anti-localization, unusual magneto-electric effects, and controlled electronic mass by an applied magnetic field, or a proximity to quantum orders.\cite{Hasanreview,Zhangreview,TSCFuKane,TISC,monopole,axion} The presence of a Dirac cone at the Fermi level without the intervention of any bulk state will also make the TI an attractive candidate to realize the fractional quantum statistics and non-Abelions.\cite{Hasanreview,Zhangreview,FuKane07,monopole,axion}. Still with all the exciting developments, it has proven to be challenging to obtain a true bulk insulator, or having a surface Dirac point at the Fermi level.

We propose here an alternative approach to design TIs by combining a set of layers of two dimensional Fermi gases (2DFGs) with Rashba-type spin-orbit coupling. This two-dimensional spin-orbit locked metallic state, in the presence of interlayer quantum tunneling, translates into a bulk insulator with $Z_2$-invariant topological properties and Dirac excitations on the surface. The idea is to grow two counter-helical Rashba-planes (dubbed `Rashba-bilayer') $-$ which hitherto imposes time-reversal invariance $-$ along the (001)-axis with an interlayer distance that enables single-electron hopping between them. With an effective Hamiltonian, we observe that above a critical number of Rashba-bilayers, about 5-6 layers for a realistic parameter choice, the non-trivial TI phase emerges. The resulting gapless single-Dirac cone has a linear slope determined by the Rashba-coupling strength and is thus externally tunable. We find that this `layer by layer' approach of Rashba-bilayers has all the known properties of the bulk TI; for example we present a direct calculation of the Chern index in the bulk to support the topological nature of the resultant state. A design principle for such Rashba bilayer with the help of a ferroelectric substrates to the 2DFG is also proposed below. With the rapid expansion of surface growth techniques like molecular beam epitaxy, one would be able to explore a large set of compounds that might not be accessible in the bulk phase and yet possess TI behaviors.

In what follows, we lay a general framework for generating and manipulating a `homemade' three dimensional TI.  In principle one can envision to use the approach we propose and take it a few steps further. For example, a heterostructure setup allows one to introduce layers of magnetism,\cite{monopole,axion} superconductivity\cite{TSCFuKane,TISC,Bi2Se3TF} and other exotic many body orders\cite{SODW} within the topological matrix. Taken together, the search for TI materials that obey several symmetry properties and inherits spin-orbit coupling can thus be replaced with `homemade' systems by generating spin-orbit coupling via external or internal electric fields, and by imposing symmetry properties via manipulating the heterostructure geometry. Such TI will be free from any particular crystal geometry studied earlier.\cite{bigraphene}

\begin{figure}
\rotatebox[origin=c]{0}{\includegraphics[width=1.0\columnwidth]{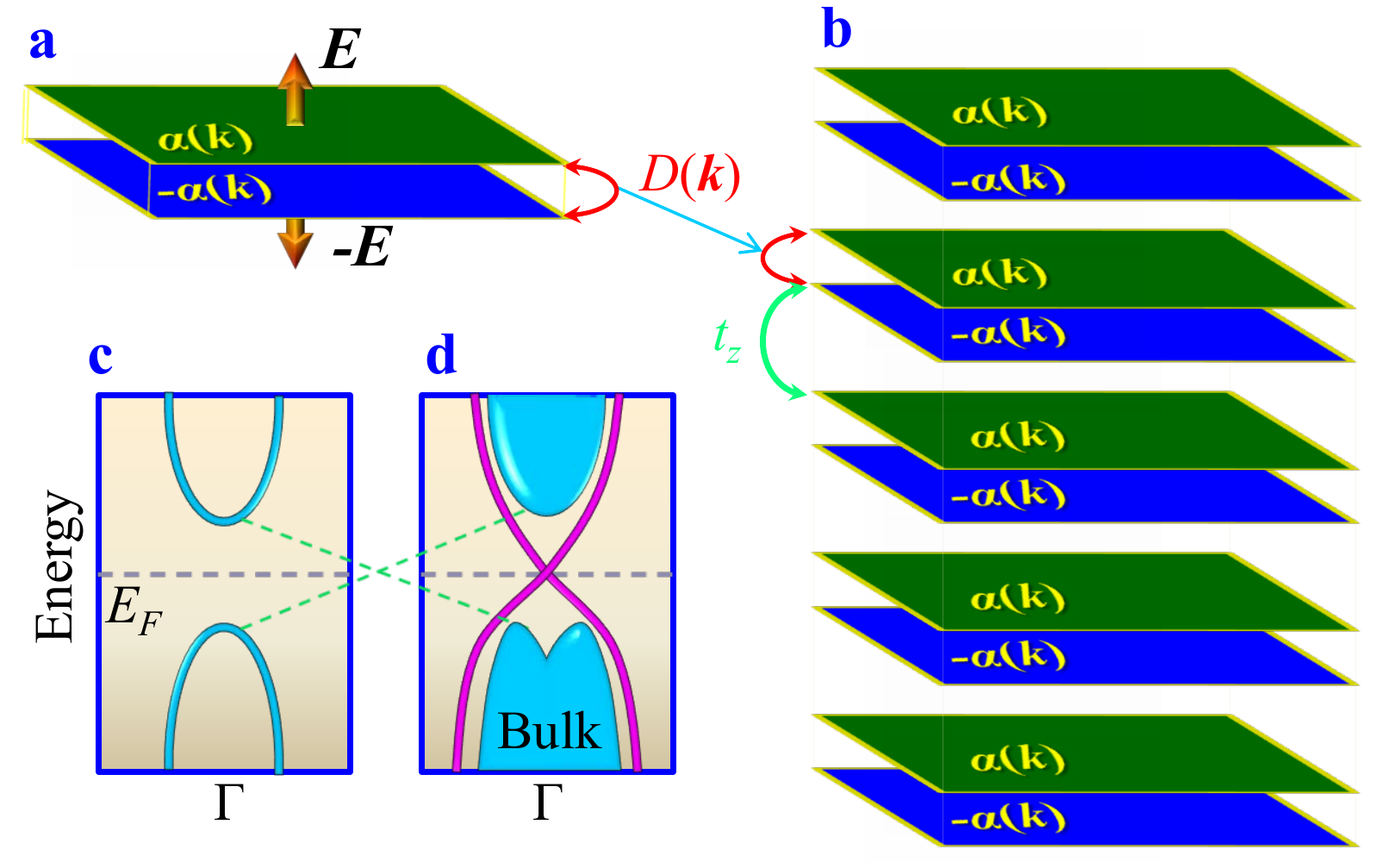}}
\caption{{\bf Rashba-bilayer of 2DFGs and its heterostructure setup.} {\bf a}, A bilayer combination of opposite Rashba-type spin-orbit coupling 2DFGs, denoted by $\alpha({\bf k})$, $-\alpha({\bf k})$, representing $h_{\rm R}^+$ and $h_{\rm R}^-$, respectively in the Hamiltonian in Eq.~1. $D({\bf k})$ gives the interlayer electron tunneling between them. {\bf b}, As grown Rashba-bilayers with finite interlayer coupling, $t_z$. {\bf c}, {\bf d}, Illustration of band dispersions for a bilayer 2DFG, and its heterostructure version, respectively. The emergence of an inverted band curvature in the valence bulk band marks the topological phase transition, as also demonstrated in first-principles bandstructure calculations,\cite{MKlintenberg} and ARPES data\cite{TIPTHasan}.} \label{setup}
\end{figure}

\vskip0.25cm\noindent
{\bf Results}\\
{\bf Rashba-bilayer heterostructure.} We start with depicting our basic idea in Fig.~\ref{setup}. The main idea relies on gluing two Rashba-type spin-orbit coupled 2DFGs with opposite signs of Rashba coupling, denoted by $\pm\alpha({\bf k})$. We take them to be close to each other such that quantum tunneling [$D({\bf k})$] couples them, as illustrated in Fig.~\ref{setup}a. Such opposite-coupled Rashba-bilayer can easily be manufactured by creating a potential gradient between two 2DFGs with the help of gating, or by inserting oppositely polarized ferroelectric substrate between them, among others. As opposed to a metallic single-Rashba 2DFG, the Rashba-bilayer opens an insulating gap, determined by the value of $D({\bf k})$, with a minimum gap at the $\Gamma$-point. Then this Rashba-bilayer setup is to be repeatedly grown along the (001)-direction with an inter-bilayer electron hopping, $t_z$, which is required to be different from $D(0)$ to eliminate the degeneracy in the band structure. Above a critical number of the Rashba-bilayers in such a heterostructure setup, a bulk topological phase transition commences. A good indicator of the `non-trivial' topology is the development of an inverted band dispersion or `dent' in the valence Fermi sea, as illustrated in Fig.~\ref{setup}d. Such an inverted band dispersion is well-established in first-principles band structure calculations,\cite{MKlintenberg} and angle-resolved photoemission spectroscopy (ARPES) measurements\cite{TIPTHasan}. Both analytically and numerically, we investigate a realistic parameter space and find that the state is a non-trivial topological state with a single Dirac cone, carrying all salient topological properties that were derived and realized earlier in bulk three-dimensional systems.\cite{Hasanreview,Zhangreview} A detailed progression of the resulting band structure is given in the Supplementary Methods.

\vskip0.25cm\noindent
{\bf Effective model for the Rashba-bilayer heterostructure.} Based on the above-proposed setup, we now derive an effective low-energy theory. In each 2DFG planes, electrons with momentum ${\bf k}$ experience an effective `anisotropic' magnetic field, induced by an electric field ${\bf E}_z$, which couples to their spin ${\bf \sigma}$ to give rise to Rashba-type spin-split electronic bands $h_{\rm R}^{\pm}=k^2/2m^* \pm \alpha_{\rm R}({\bf k}\times{\bf \sigma})$. Here $\alpha_{\rm R}$ is the Rashba-coupling strength, controlled by the external or internal electric field, and $m^*$ is the effective mass of electrons. ${\bf k}$ is defined in the 2D plane, and correspondingly ${\bf \sigma}$ stands for Pauli matrices in the spin-subspace. We assume two such counter-propagating helical 2DFGs, $h_{\rm R}^+$ and $h_{\rm R}^-$, are grown close to each other. Because of wavefunction overlap, finite quantum tunneling $D({\bf k})$ between two 2DFGs is active. The next step in this setup is to grow these Rashba bilayers along $z-$axis with a spin-conserving electron hopping between them, characterized by $t_z$. Therefore, the general form of the effective Hamiltonian for $N$ semi-infinite layers within an open-boundary condition can be expressed as
\begin{eqnarray}
&&H=
\left(
\begin{array}{cccccc}\
h_{\rm R}^+     & D  & {\bf 0} & {\bf 0} & \ldots \\
D  & h_{\rm R}^-     & T       & {\bf 0} & \ldots \\
{\bf 0}         & T               & h_{\rm R}^+     & D & \ldots   \\
{\bf 0}         & {\bf 0}         & D  & h_{\rm R}^-     &  \ldots      \\
\vdots & \vdots & \vdots & \vdots & \ddots\\
\end{array}
\right)
\label{Ham}
\end{eqnarray}
Each term in the above Hamiltonian is a $2\times2$ matrix. To keep the formalism general and readily tunable, we allow for `anisotropic' hopping between $h_{\rm R}^+$ and $h_{\rm R}^-$ as $D({\bf k})=(D_0+Mk^2)\mathrm{I}_{2\times2}$, which can also be thought of as Dirac mass and Newtonian mass terms, respectively, due to their impacts on the band structure obtained. The tunneling between two adjacent bilayers is $T=t_z{\mathrm{I}}_{2\times2}$. ${\bf 0}$ is a $2\times2$ zero matrix, and $\mathrm{I}_{2\times2}$ is the identity matrix. Time-reversal invariance of the above Hamiltonian consequently emerges due to the fact that $h_{\rm R}^{-\dag}(-{\bf k})=h_{\rm R}^+({\bf k})$, which is an essential criterion for the formation of helical edge states with a Dirac point, endowing the system to the $Z_2$ classification.\cite{FuKane07,Bi2Se3SCZhang,RoyZ2,MooreZ2,Schnyder,Zaanen}

The values of parameters $m^*$ and $\alpha_{\rm R}$ depend on the type of Rashba-setup under consideration, whereas those for the expansion parameters $D$, $M$ and $t_z$ are controllable mainly by the details of the heterostructure. An elaborated band structure progression as a function of these parameters is given in Supplementary Fig.~S2. Here we discuss the emergence of a topological bulk insulator with helical surface state for a representative value of $m^*=0.1$~eV$^{-1}$\AA$^{-2}$, and $\alpha_{\rm R}=1.5$~eV\AA. Such a value of the Rashba parameter is readily achievable in Bi-based surface and quantum-well states, as well as in BiTeI bulk systems.\cite{BiTeI} The Dirac mass has the same meaning as used for 2D quantum spin Hall (QSH) insulator,\cite{HgTeSCZhang} and 3D TIs,\cite{Bi2Se3SCZhang} in that it is responsible for opening a band gap between the two adjacent time-reversal terms $h_{\rm R}^+$ and $h_{\rm R}^-$ at the $\Gamma$-point inside the bulk. When several slabs of Rashba-bilayers are glued together with $t_z$ tunneling, the insulating gap persists in every layers except on the boundary. Two helical edge states develop at each boundary and meet at a gapless Dirac cone above a critical value of the number of layers, $N_{\rm c}$.

The value of $N_{\rm c}$ depends only on the Dirac mass $D_0$ and on the interlayer tunneling $t_z$, but not on other terms, and thus is controllable solely by the `layer-by-layer' deposition technique. We illustrate this case for a representative value of $D_0=-25$~meV, $t_z=-0.2$~eV in Fig.~\ref{band}. Above three or more Rashba-bilayers, we obtain, both numerically and analytically (see Supplementary Methods), that an inverted bulk insulating gap develops in the interior Rashba-bilayers. This reflects in the transformation of the quantum-well-like states of the interior slabs into an inverted dispersion resembling a `dent' shape in the filled Fermi sea. We recall that this `dent'-like inverted valence band is a critical signature of the non-trivial topological phase, and has been consistently obtained in first-principles calculations,\cite{MKlintenberg} and has also been observed in 3D TI materials.\cite{TIPTHasan}

To further quantify the `strong' topological phase transition as a function of the number of layers, we compute the topological invariants $\nu$ (or an axion angle parameter $\theta$) of the Hamiltonian in Eq.~(\ref{Ham}). Since the Hamiltonian is invariant under inversion symmetry, we can derive the essential parity operator for each pair of Kramers degenerate valence bands from the constraint $PH(k)P^{-1} = H(-k)$, with $P=\sigma_z\otimes I_{2\times 2}$, for each Rashba-bilayer. Thus we evaluate the topological quantum index as\cite{TBTI}
\begin{eqnarray}
(-1)^{\nu}=\prod_{i}\prod_{l=1}^N\langle \phi_l|P|\phi_l\rangle,
\label{parity}
\end{eqnarray}
where $i$ stands for the time-reversal high-symmetry points on the Brillouin zone, and $\phi_l$ are the eigenvectors for each layer $l$. For the parameter set used, we indeed find that a parity inversion occurs at $N_{\rm c}$, which, according to the $Z_2$ topological criterion,\cite{FuKane07,Bi2Se3SCZhang,RoyZ2,MooreZ2} marks the emergence of the topological insulating phase at this critical value of $N$. Above this critical thickness, the parity value stabilizes to the non-trivial TI.\cite{parityoscillation}

A consequence of the $Z_2$ TI is the presence of a spin-polarized Dirac cone at the surface. Despite the emergence of a `non-trivial' topological phase above three Rashba bilayers, a surface gap persists due to finite quantum tunneling between the two edges, and the massless Dirac point appears above six layers for this parameter choice. Such a formation of massive Dirac quasiparticles below a critical value of quintuple layers is observed in Bi$_2$Se$_3$ thin films,\cite{Bi2Se3TF} and also in doped bulk TlBi(S$_{1-x}$Se$_x$)$_2$\cite{TIPTHasan,Bi2Se3doped}.

We tabulate the values of bulk and surface gaps as a function of various tuning parameters in Fig.~\ref{gap}. As expected the electron's mass $m^*$ and Rashba-coupling $\alpha_{\rm R}$ do not have any significant effect on the gap values, and thus give us an alternative approach for generating bulk TIs beyond atomistic spin-orbit coupling and heavy electron mass. The Dirac mass $D$ and interlayer tunneling provide the ``tuning knob'' for engineering the surface and bulk gaps, which are readily tunable via heterostructure details. In Supplementary Fig.~S2, we show the corresponding energy dispersions for a large range of realistic parameters.

\vskip0.25cm\noindent
{\bf Surface Dirac cone properties.} We can formulate the Hamiltonian for the the edge state by using the theory of invariants. The time-reversal symmetry imposes that a spin-up state at momentum ${\bf k}$ must be entangled to a spin-down state at $-{\bf k}$ via spin-orbit coupling. As a consequence of this, according to the Fermion doubling theorem, a gapless Dirac point is guaranteed at the $\Gamma$-point. In what follows, two counter-propagating surface states form with one spin state from the upper Rashba-term ($h_{\rm R}^+$), and an opposite spin state from the lower Rashba term ($h_{\rm R}^-$), which are related by time-reversal operation. By projecting the full Hamiltonian in Eq.~1 onto a single slab, we can write down the surface Hamiltonian in this basis to leading order in $k$ as
\begin{eqnarray}
H_{\rm surf}= \alpha_{\rm R}(\sigma^xk_y - \sigma^yk_x).
\label{Hsurf}
\end{eqnarray}
Note that the similar surface dispersion is also obtained for 3D TIs in Refs.~\cite{FuKane07,Bi2Se3SCZhang}, except that here its slope is solely determined by the Rashba-coupling strength. Therefore, the velocity of the Dirac fermions turns out to be $v=\alpha_R/\hbar$, and externally tunable. For values of the Rashba-coupling constant as large as 3.8~eV$\AA$, achieved to date in bulk BiTeI,\cite{BiTeI} we get $v\simeq5.8\times10^5$~ms$^{-1}$, which is several order of magnitude larger than the value achieved so far in 3D TIs.\cite{Hasanexp,Ong} The energy scales above which the slope of the surface state deviates from a linear-in-energy to a power-law behavior depends on the value of mass term $D(0)$. For the parameter set used above, we obtain a linear dispersion expanding about $0.3$~eV on both sides of the Dirac point, which is also tunable (see Supplementary Fig.~S2).

The spin-polarized surface states can be imaged directly by ARPES and tunneling spectroscopies, as well as via transport measurements. An interesting transport property of Dirac fermions is the quantum Hall effect which can be achieved by inserting a magnetic layer in the heterostructure next to the boundary, so that it breaks time-reversal symmetry on the surface via the proximity effect, but not in the bulk.\cite{FuKane07,monopole} We can express the surface and magnetic layer Hamiltonian in a combined `d'-vector form as ${\bf d}\cdot{\bf \sigma}$, where the time-reversal invariant components are $d_x=-\alpha_{\rm R} k_x$ and $d_y=\alpha_{\rm R} k_y$, and time-reversal breaking $d_z$ gives the Zeeman energy splitting due to induced magnetization. Without the presence of a $d_z$ term, the winding number of the upper and lower chiral states, defined as $C=\frac{1}{4\pi^2}\int dk_x\int dk_y \hat{\bf d}\cdot(\partial_x \hat{\bf d}\times\partial_y \hat{\bf d})$ (the notation `hat' represents corresponding unit vector) where the integration is performed over a closed loop around the Dirac cone, cancel each other. However, when a broken time-reversal symmetry is imposed, $C$ becomes equal to $\pm\sigma_z$, where $\sigma_z$ is the component of the spin along the magnetic field orientation. In this case, quantized spin-Hall conductance becomes fractional in units of $e^2/h$ as $\sigma_H=\mp(|\sigma_z|/2)(e^2/h)$. For a fully polarized spin-configuration along the magnetic field, the above equation generates a half-integer anomalous quantum-Hall effect (QHE). The fractional or half-integer QHE is a trademark feature of Dirac systems,\cite{Hasanreview,Zhangreview,FuKane07,Ong,graphene} and can be used as a test of our proposal. An interesting consequence of the half-integer Landau level is that as the time-reversal breaking mass term, $d_z$, approaches zero, the counter-propagating Landau levels move to zero energy, and the system can act as protected quantum Hall ``insulator'' despite the presence of energy states at the Fermi level\cite{Jackiw}. This unique scenario has been discussed theoretically for TIs, but never been realized experimentally due to the absence of an isolated Dirac point at $E_F$,\cite{Ong} however, it can be realized in the present heterostructure setup.

\vskip0.25cm\noindent
{\bf Design principles for Rashba bilayers} Finally, we propose two representative design principles for the counter-polarized Rashba-type spin-orbit coupling bilayers in Fig.~\ref{fig4}.  Of course, the materials fabrication of them is not limited to these two design methods, and one can envision methods by exploiting our general idea of invoking a polarized medium inside the Rashba-bilayer. Here, we suggest to use the oppositely polarized ferroelectric materials as the substrate to two adjacent 2DFGs. The idea is to utilize the uniform polarization of a ferroelectric material to tune the Rashba-type spin-orbit coupling at the interface.\cite{EFRashba} We think of systems such as ferroelectric polymers and BiFeO$_3$/La$_{1-x}$Sr$_x$MnO$_3$ (BFO/LSMO) superlattice as possible candidates for such substrates which come with an additional benefit that they are highly strain-free . Another crucial advantage of this design principle is that the ferroelectricity is intrinsic in these these materials, and it is generated by the inversion symmetry breaking in the bulk. Thus one can easily generate oppositely aligned ferroelectric polarization on both edges by creating a reverse inversion symmetry breaking with respect to a common mirror {\em ab}-plane between them.

In our first materials specific set up, we propose to use a copolymer of vinylidene fluoride with trifluoroethylene, P(VDF-TrFE), consisting of -((-CF$_2$-CH$_2$)$_x$-(-CF$_2$-CHF-)$_{1-x}$)$_n$- chains, controlled by the C-C bonds between fluorine pairs,\cite{polymer} as shown in Fig.~\ref{fig4}a1. This polymer exhibits spontaneous polarization as large as 0.1Cm$^{-2}$, and a hysteresis loop for switching polarization even in two monolayer films.\cite{polymerpolarization} The polarization is highly homogeneous and tunable. The electric field is generated from Fluorine to Hydrogen ions due to the lack of inversion symmetry in the polymer as illustrated in Fig.~\ref{fig4}a1. This electric field can be exploited in two ways to generate a Rashba-bilayer. In Fig.~\ref{fig4}a2, we show that two P(VDF-TrFE) polymers can be attached back to back (setting either Fluorine or Hydrogen layer as a mirror plane). If a 2DFG with heavy elements such as Bi is attached on both sides of this polymer, they may generate opposite Rashba-type spin-orbit couplings. In this case, since the mirror plane intrinsically maintains a constant charge character, two opposite electric fields are unlikely to annihilate each other. On the other hand, since the dc conductivity is quite large in this film,\cite{polymerconductance} a finite electron tunneling between the two Rashba-layers can turn on due to the wavefunction overlap. This setup will then work as Rashba-bilayers of interest.

In Fig.~\ref{fig4}a3, we propose a second Rashba-bilayer design scenario using this polymer. Here two polymers can be attached from the top and the bottom sides of the 2DFG-bilayers, enabling an opposite Rashba spin-orbit coupling. In between them, one needs to add a metallic substrate which has a polar surface (having opposite polarity on both sides may make a metal even more suitable here). Since this polymer is very flexible, we can expect to obtain a strain-free, homogeneous counter helical Rashba-type spin-orbit coupling on two adjacent 2DFGs.

In the second setup, we propose to use a ferroelectric BFO/LSMO superlattice. Here the interfacial valence mismatch between BFO and LSMO (without any significant structural mismatch) influences the electrostatic potential step across the interface, which manifests itself as the bias-voltage in the ferroelectric hysteresis loops. As shown in Figs.~\ref{fig4}b1 and \ref{fig4}b2, in this material, the polarization is reversible depending on how LSMO layer is attached to the BFO layer.\cite{BFOLSMO} We can exploit these properties to generate two Bi-layers with opposite polarizations, as demonstrated in Figs.~\ref{fig4}b3 and \ref{fig4}b4. The added benefit here is the strong spin-orbit coupling of Bi atoms, which will thus provide an opposite Rashba-type spin-orbit coupling in two adjacent Bi-layers. Since LSMO can be doped easily from paramagnetic to semimetal to (trivial topological) insulator phase, the inter-layer hopping is easily tunable in this setup. Since the BFO/LSMO interface heterostructures are readily grown, our proposed BFO/LSMO/BFO structure can also be expected to be achievable in the same way. Despite the fact that the magnetic moment in BFO is mainly on the Fe sites, it may be suggestive to avoid the magnetic ground state of BFO to achieve a full spin-polarization of the Rashba-type spin-orbit effect. The possibility of acquiring a ferroelectricity controlled Rashba spin-orbit coupling in BaTiO$_3$ as demonstrated by first-principle calculation\cite{EFRashba} also makes it a suitable candidate for generating a topological insulator in the superlattice structure with LSMO. Finally, we note that the infinitely adaptive superlattice phase of single layers of Bi and Bi$_2$Se$_3$, as opposed to bilayers of each of them used in Ref.~\cite{Bi2Bi2Se3} which is a topological semimetal, can become a topological insulator.

\vskip0.25cm\noindent
{\bf Discussion}\\
Our proposal clearly overcomes the limitation of searching for suitable combinations of crystal geometry, and inherent wave function symmetries within a bulk system to obtain TI. The present formalism is free from any particular crystal geometry, and thus provides a widespread playground for engineering `homemade' TIs with surface Dirac state properties. Another advantage of the present case is that here the edge state is isolated from the bulk states, instead of being buried inside the bulk Fermi sea, which has so far significantly limited the usage of existing 3D TIs. The heterostructure, with depositing capability of one atomic layer at a time, can also easily accommodate magnetic and superconducting layers, beyond the conventional doping or proximity effects. It offers a freedom of bringing the desired components to the relevant part of a nanostructure by demand. This ability thus will promote a unambiguous detection of non-Abelian particles\cite{TSCFuKane,monopole,axion} and anomalous Hall effect. Given that the electron interaction is strong and tunable in 2DFG,\cite{SODW} novel broken symmetry phases are easy to yield in this heterostructure setup than in a weakly correlated 3D TI.

\clearpage
\newpage

 \vskip0.25cm\noindent
{\bf Acknowledgements}\\
We are indebted to S. Basak, Q. Jia, H. Lin, J. Halardsen, and A. M. Black-Schaffer for numerous discussions. We also thank Paul Ruden for critical reading of our manuscript. This work is supported by by Los Alamos National Laboratory under LDRD, by the US Department of Energy under BES, and benefited from the allocation of supercomputer time at NERSC. Work at Nordita is supported by VR 621-2012-2983 and by ERC DM 321031.

 \vskip0.25cm\noindent
{\bf Author contributions}\\
The present research stemmed from fruitful discussions between the authors. Both authors contributed to the writing of the manuscript.

 \vskip0.25cm\noindent
{\bf Additional information}\\
{\bf Competing financial interests:} The authors declare no competing financial interests.\\
{\bf Reprint and permission} information is available at \\
{\bf How to cite this article}

\clearpage
\newpage

%

%
\begin{figure}
\rotatebox[origin=c]{0}{\includegraphics[width=1.0\columnwidth]{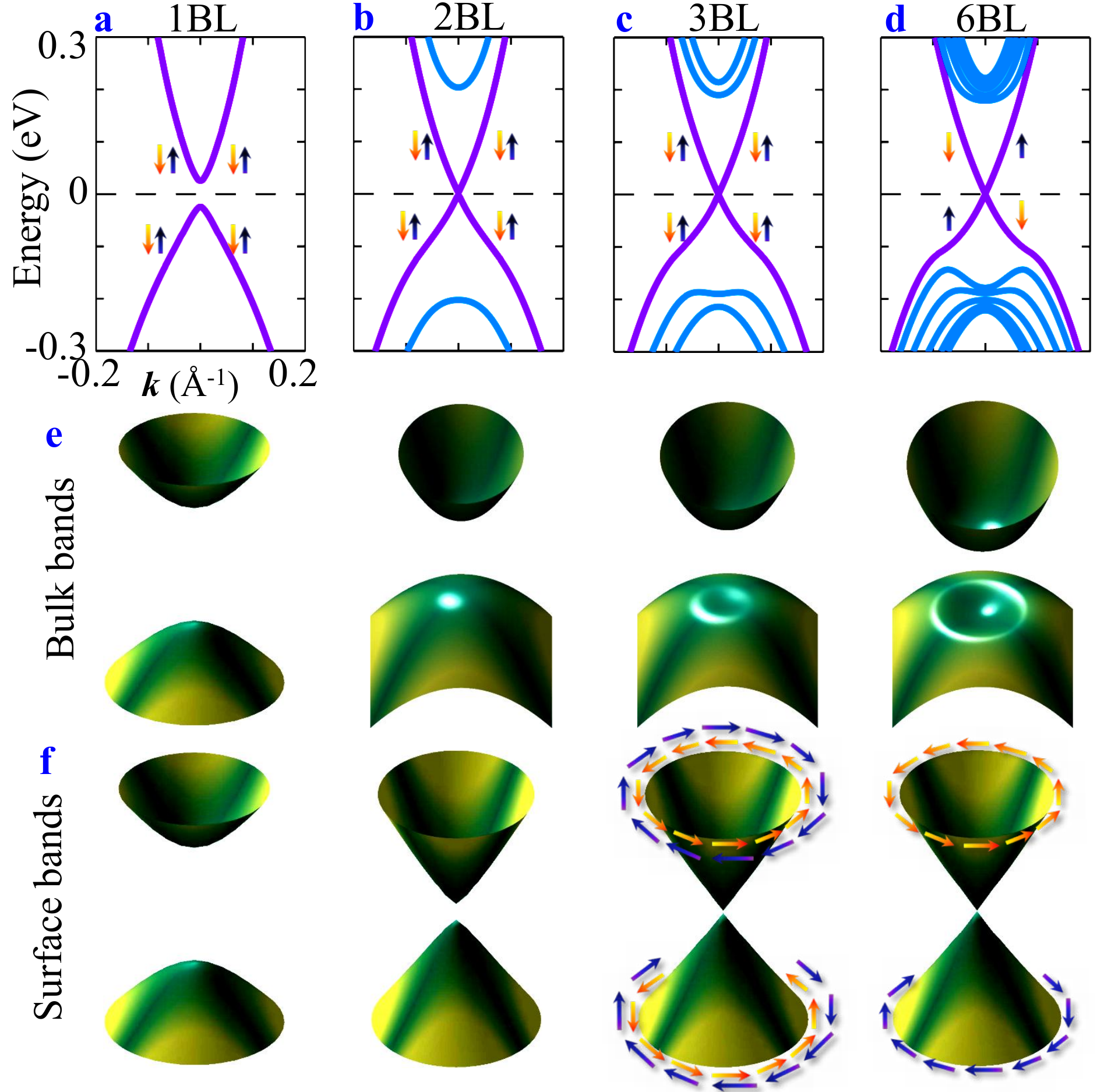}}
\caption{{\bf Band progression and formation of gapless surface states.} {\bf a}-{\bf d}, Evolution of band dispersion for a single (1BL) to six (6BL) of Rashba-bilayers. {\bf e}-{\bf f}, Corresponding 3D view of the band dispersions in the $k_x,~k_y$-plane for the bulk (top panel) and surface (bottom panel) bands, respectively. The parameter set for this calculation is given in the text. In going from two to three bilayers, the bulk valence band topology reveals the emergence of an inverted shape [see also {\bf c} and corresponding {\bf f}], which indicates the topological phase transition from a trivial to the non-trivial phase. However, it takes about six bilayers to turn off the inter-edge tunneling to commence a gapless Dirac cone at the surface. Arrow dictates the spin orientation. The definite spin-chirality of the gapless Dirac cone is illustrated by counter-rotating arrows in {\bf f}. Green (min) to blue (max) colormap gives projected spin-orbit locking eigenstates.} \label{band}
\end{figure}
\begin{figure}
\rotatebox[origin=c]{0}{\includegraphics[width=1.0\columnwidth]{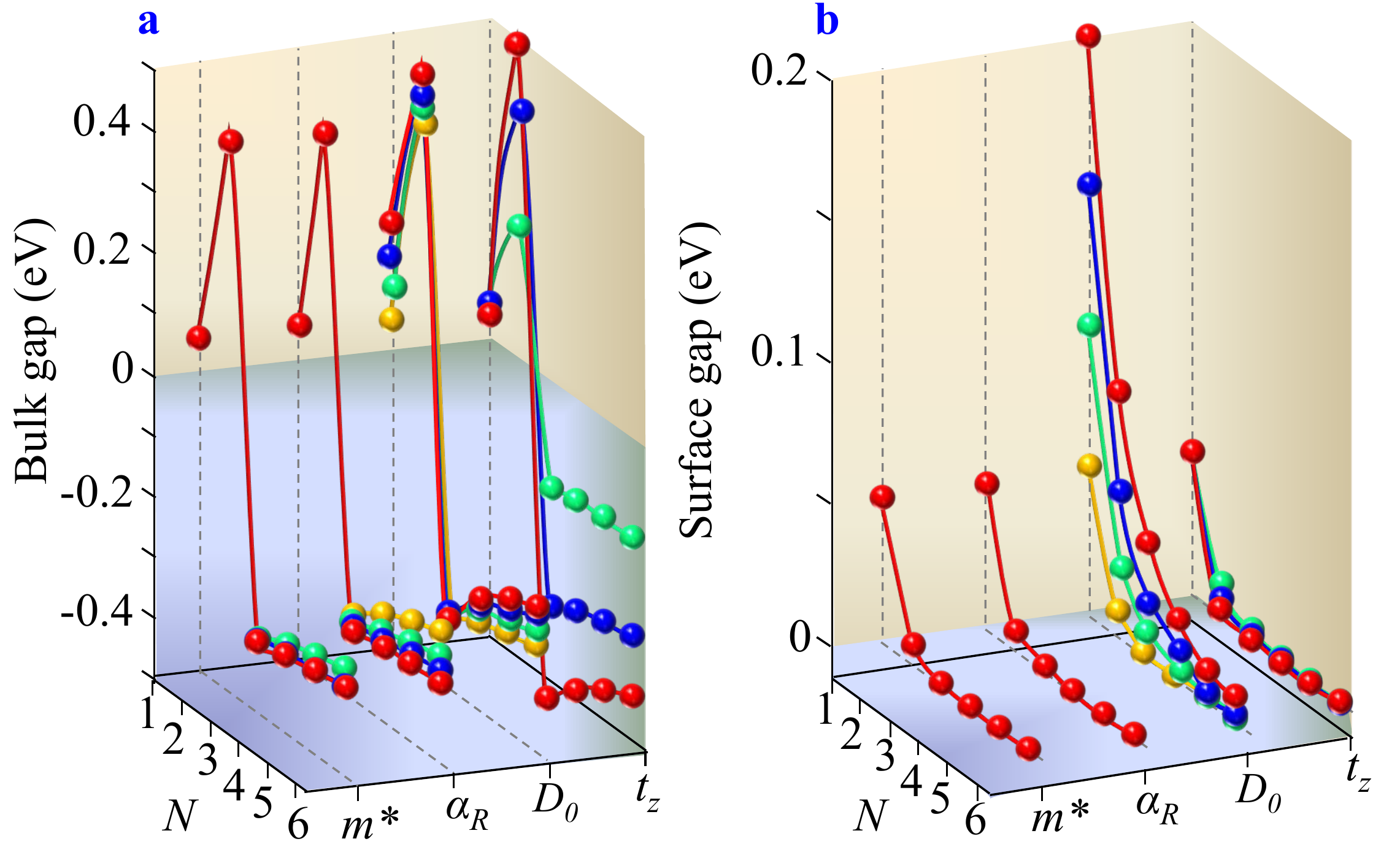}}
\caption{{\bf Bulk and surface gaps in a realistic parameter space.} {\bf a}, The critical number of layers, $N_c$, at which an inverted bulk gap opens. The bulk band gap jumps from $D(0)$ at $N$=1 (equivalent to $t_z$=0) to a large positive value, and then consistently becomes negative for $N\geq 3$, constrained by the Hamiltonian (see Supplementary Fig.~S2). For each parameter set, the corresponding other parameters are set to be constant to the values mentioned in the main text. The varying parameters are: $m^*$=1 (green), 2.5 (blue), 5 (red) in eV$^{-1}$\AA$^{-2}$. $\alpha_{\rm R}$=1 (orange), 1.5 (green), 2 (blue), and 3 (red) in eV\AA which are attainable in existing systems.\cite{BiTeI} $D_0$=-25 (orange), -50 (green), -75 (blue), and -100 (red) in meV, and $t_z$=-100 (green), -200 (blue) and -500 (red) meV. Since $Mk^2=0$ at the $\Gamma$-point, the parameter $M$ does not have any effect on the direct gap structure. {\bf b}, Corresponding surface gaps at the $\Gamma$-point. Clearly, the gapless surface Dirac cone formation depends on two parameters, the interlayer hybridization $t_z$, and strongly on the Dirac mass $D$. The parameter values here are same as in {\bf a}.} \label{gap}
\end{figure}

\begin{figure}
\rotatebox[origin=c]{0}{\includegraphics[width=1.0\columnwidth]{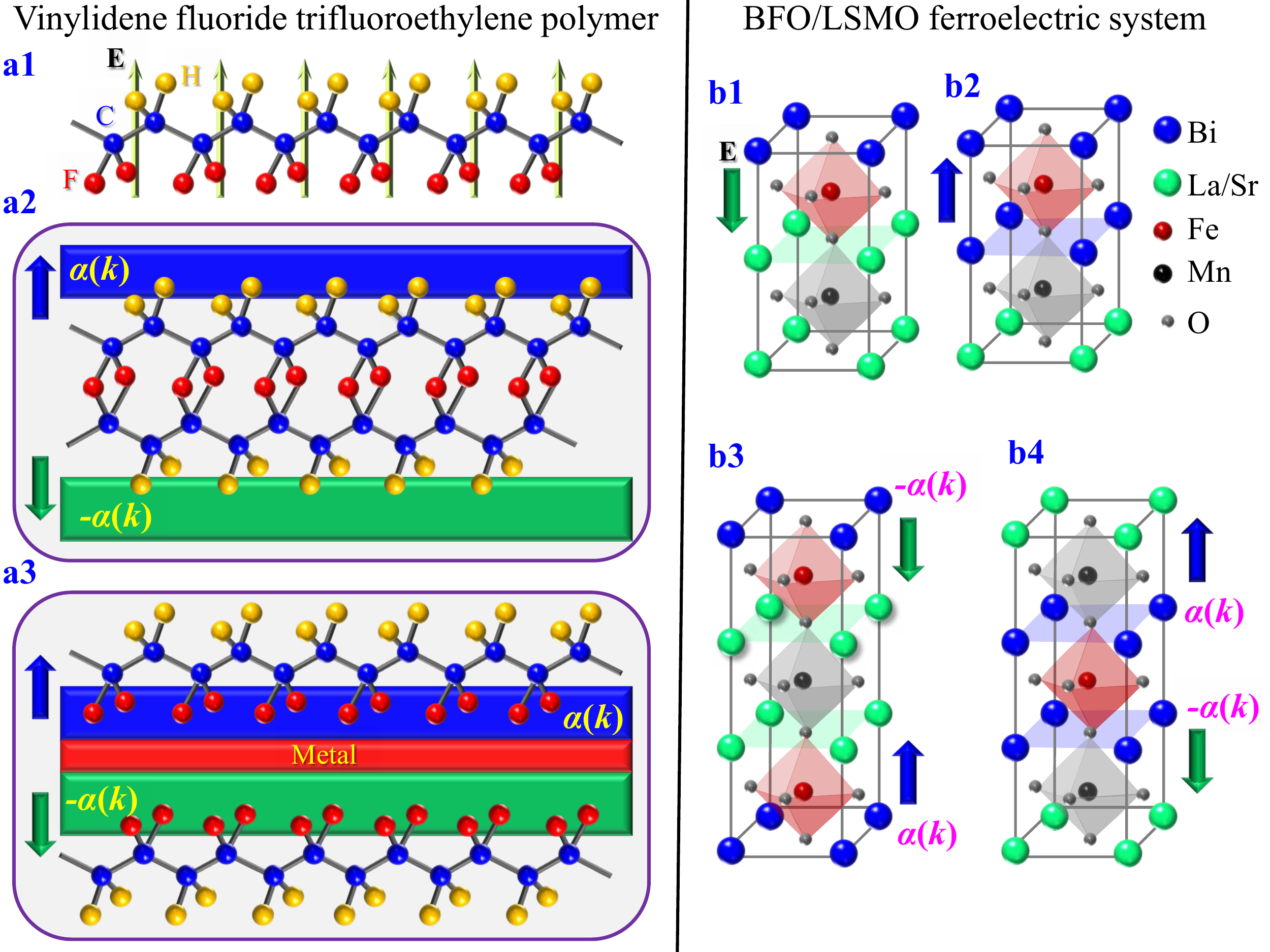}}
\caption{{\bf Design principles for counter helical Rashba-bilayer via ferroelectric substrates.} {\bf a1}, In the left panel, we show the structure of a copolymer of vinylidene fluoride with trifluoroethylene, P(VDF-TrFE) which exhibits ferroelectricity even in its two-dimensional films.\cite{polymer} Two oppositely oriented such films can be used to generate oppositely polarized Rashba spin-orbit couplings in the adjacent layers of 2DFGs. Two possible design principles using this polymer are presented in {\bf a2} and {\bf a3}. Blue and green arrows depict opposite directions of electric field. $\pm\alpha({\bf k})$ distinguish two 2DFGs with opposite Rashba-type spin-orbit coupling. {\bf b1}-{\bf b4}, In the right panel, we demonstrate the design principles of Rashba-bilayer by exploiting the ferroelectric BFO/LSMO superlattice.\cite{BFOLSMO} Here the existing Bi-layers are expected to possess an oppositely polarized Rashba spin-orbit coupling via manipulating the direction of the inversion symmetry breaking inside them.}\label{fig4}
\end{figure}

\clearpage
\newpage

\begin{center}

\large{\bf Supplementary Information for ``Engineering three-dimensional topological insulators in Rashba-type spin-orbit coupled heterostructures"}\\

\vskip0.5cm
Tanmoy Das, A. V. Balatsky \\

\vskip1cm

\end{center}



\noindent
{\Large{\bf Suplementary Figures}}\\\\



\begin{figure}[h]
\rotatebox[origin=c]{0}{\includegraphics[width=.9\columnwidth]{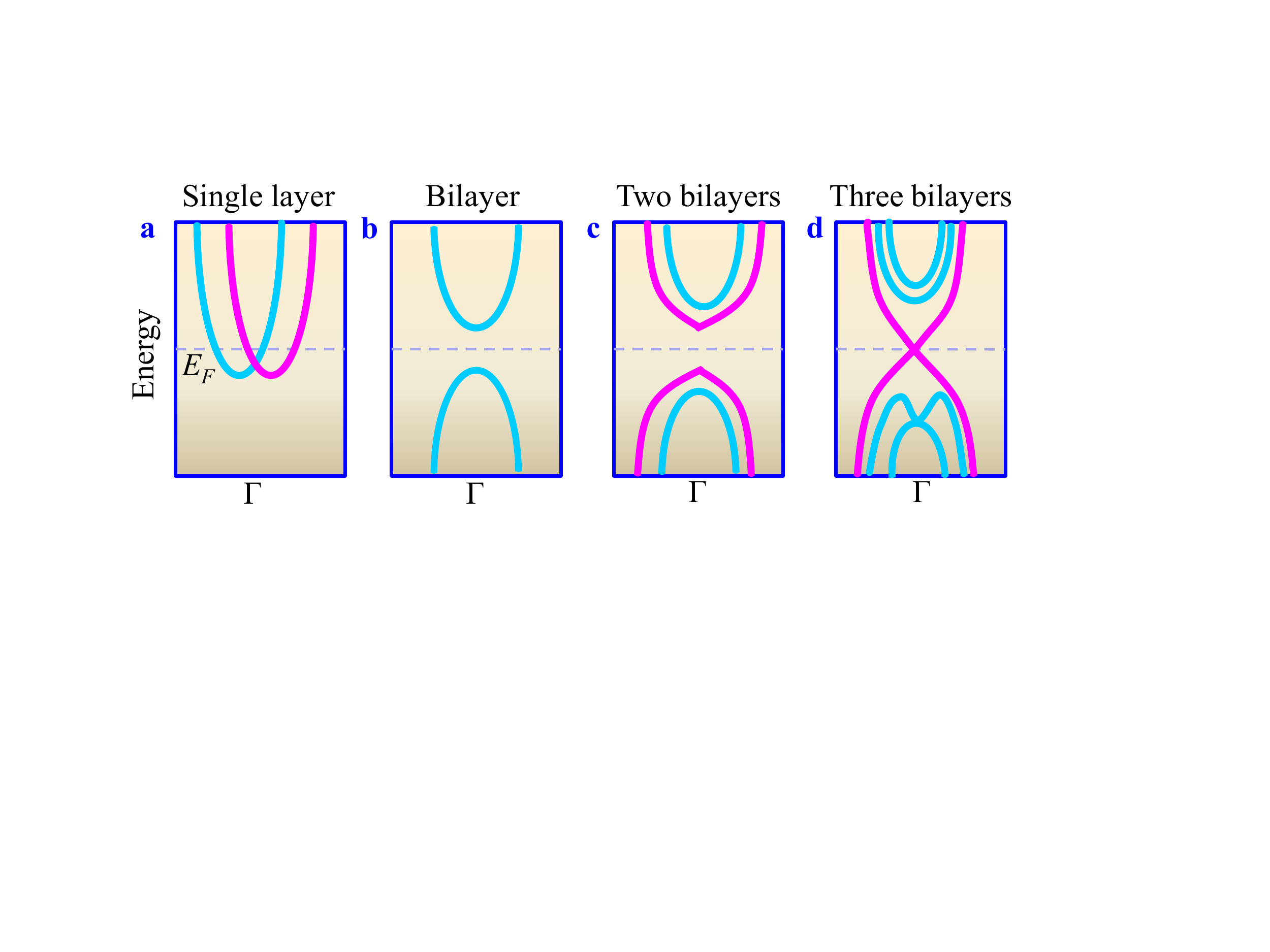}}
\vspace{12pt}\\\parbox{\textwidth}{
{\bf Supplementary Figure S1: Band progression from a single to a bilayer to a heterostructure of Rashba-type spin-orbit coupling 2DFGs.} Schematic drawings of the band curvatures expected for a single- to a bi- to a multiple layers of 2DFGs.}
\end{figure}

\clearpage
\newpage
\begin{figure}[h]
\rotatebox[origin=c]{0}{\includegraphics[width=1.0\columnwidth]{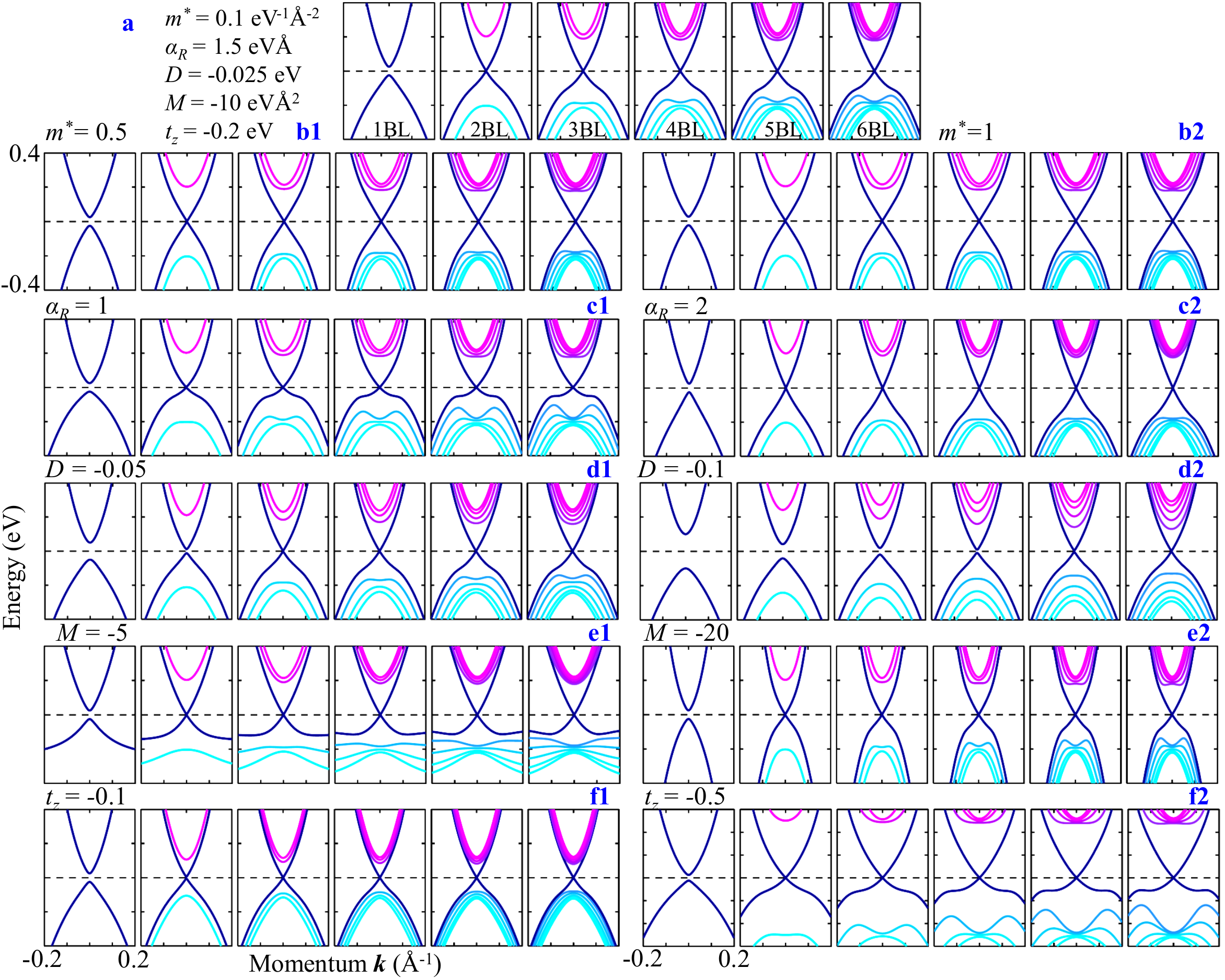}}
\vspace{12pt}\\
\parbox{\textwidth}{
{\bf Supplementary Figure S2: Variations of the dispersion relation for a wide range of parameter space.}
{\bf a}, The band dispersion for a bilayer 2DFG to six layers setup as shown in Fig.~1 of the main text. In the rest of the figures, only a single parameter is tuned (in each row), while keeping others parameters same to that in {\bf a}. In each figure, the horizontal axis spans within the $k_x=$ -0.2 to 0.2 \AA$^{-1}$ range at $k_y=0$ and the vertical axis runs from $E=$-0.4 to 0.4~eV with Fermi level ($E_F$) lying at the center (black dashed line), except in {\bf f2}, where the energy scale runs from -0.6 to 0.6~eV. The value of $N_{\rm c}$ for the formation of gapless surface state depends only on $t_z$ and $D_0$, whereas the slope of the surface state is controlled by the Rashba-type SOC $\alpha_{\rm R}$ (compare {\bf b}, {\bf c1}, and {\bf c2}), and the span of the linearly dispersing surface state depends on the bulk insulating gap, as well as on the Newtonian mass $M$, as expected form Eq.~2 of the surface Hamiltonian.}
\label{suppband}
\end{figure}




\clearpage
\newpage

\vskip0.25cm\noindent
{\Large{\bf Suplementary Methods}}\\

\noindent
$\bullet$~{\bf Band dispersion for a Rashba bilayer}\\

We provide further insights into the emergence of the TI from an analytical description. The Rashba-type SOC splits a non-interacting band into two in-plane spin-polarized bands. To keep the formalism general and concrete, we take a 2DFG, with a quadratic band, $k^2/2m^*$, where $m^*$ is the effective mass of electrons. Under Rashba-type SOC,[41] the split bands are obtained by solving $h_{\rm R}^{\pm}=k^2/2m^*I_{2\times 2}\pm\alpha_{\rm R}(k_y\sigma_x+k_x\sigma_y)$, where $\alpha_{\rm R}$ is the Rashba-type SOC. For simplicity, we define the second term as $\alpha({\bf k})$. Next we create a Rashba bilayer with two counter-propagating helical 2DFGs, $h_{\rm R}^+$ and $h_{\rm R}^-$ in which a finite quantum tunneling is turned on, as denoted by $D({\bf k})$. The Hamiltonian for the Rashba bilayer can be written explicitly as
\begin{eqnarray}
~~~~~~~~~~
&&H=
\left(
\begin{array}{cccc}\
k^2/2m^* & \alpha({\bf k}) & D({\bf k}) & 0  \\
\alpha^{\dag}({\bf k}) & k^2/2m^* & 0 &  D({\bf k}) \\
 D({\bf k}) & 0 & k^2/2m^* & -\alpha({\bf k}) \\
 0 & D({\bf k}) & -\alpha^{\dag}({\bf k}) & k^2/2m^*\\
\end{array}
\right).
\hspace{145pt}{\rm (S1)} \nonumber
\label{biRashba}
\end{eqnarray}

The resulting eigenvalues are two-fold degenerate: $E^{\pm}({\bf k})=k^2/2m^*\pm\sqrt{\alpha^2({\bf k}) + D^2({\bf k})}$. This specific bilayer setup helps generate two upward and downward dispersing bands, with a direct gap between them defined by $D({\bf k})$ at $\Gamma$-point (since $\alpha(0)=0$), see Supplementary Fig.~S1b. $D({\bf k})$ shall be even under inversion, e.g. $D({\bf k})=D_0+Mk^2$, which destroys the spin-polarization, implying that the gap between the band bands is charge insulating.

\vskip0.5cm\noindent
$\bullet$~{\bf Band dispersions for two Rashba-bilayers}\\

As depicted in Fig. 1 of the main text, we now take two Rashba bilayers, grown along the (001)-direction such that a quantum tunneling, say $t_z$, between the adjacent layers becomes active. The explicit form of the Hamiltonian for this case is
\begin{eqnarray}
&&H=
\left(
\begin{array}{cccccccc}\
k^2/2m^* & \alpha({\bf k}) & D({\bf k}) & 0 & 0 & 0 & 0& 0 \\
\alpha^{\dag}({\bf k}) & k^2/2m^* & 0 &  D({\bf k}) & 0 & 0 & 0& 0\\
D({\bf k}) & 0 & k^2/2m^* & -\alpha({\bf k}) & t_z & 0 & 0& 0\\
0 & D({\bf k}) & -\alpha^{\dag}({\bf k}) & k^2/2m^* & 0 & t_z & 0& 0\\
0 & 0 & t_z& 0 & k^2/2m^* & \alpha({\bf k}) & D({\bf k}) & 0  \\
0 & 0 & 0& t_z &\alpha^{\dag}({\bf k}) & k^2/2m^* & 0 &  D({\bf k}) \\
0 & 0 & 0& 0 & D({\bf k}) & 0 & k^2/2m^* & -\alpha({\bf k}) \\
0 & 0 & 0& 0 & 0 & D({\bf k}) & -\alpha^{\dag}({\bf k}) & k^2/2m^*\\
\end{array}
\right).~{\rm (S2)} \nonumber
\label{biRashba2}
\end{eqnarray}
The eigenvalues of Supplementary Eq.~(S2) are given by
\begin{eqnarray}
~~~~~~~
E^{\pm}_{\pm}({\bf k}) = \frac{k^2}{2m^*} \pm \frac{1}{2}\left[2t_z^2+4\left(\alpha^2({\bf k}) + D^2({\bf k})\right)\pm 2t_z \sqrt{t_z^2+4D^2({\bf k})}\right]^{1/2}.\hspace{85pt}{\rm (S3)} \nonumber
\label{biRashbaE2}
\end{eqnarray}
The energy dispersions imply that whereas a large bulk gap opens between $E^{+}_+$ and $E^-_-$, the gap between the lowest energy states $E^+_-$ and $E^-_+$ decreases from their single bilayer value of $D_0$. Although, mathematically, there exists a solution for the values of $D_0$ and $t_z$ at which the surface gap can be closed, but there is no clear evidence of the bulk band party inversion for this Hamiltonian, and thus the resulting solution remains to be a trivial topological phase. The absence of the band inversion is also evident in the parity calculation and the parabolic shape of the band dispersion as presented in the main text.

\vskip0.5cm\noindent
$\bullet$~{\bf Band dispersions for three Rashba-bilayers}\\

With three or more bilayers, coupled by finite quantum tunneling, the bulk systems begin to form in which the interior bilayer(s) acts as a bulk lattice where the edge layers give rise to the surface states. We show here that a bulk band inversion occurs for three bilayers, and thus promotes a non-trivial topological insulating phase. However, to get rid of the electron hopping from the two edges, or in other words, to close the surface gap, one requires a higher number of bilayers.

The analytical solution gradually becomes complicated and long with further increase of bilayers. However we can gain some insights into how an inversion occurs in three bilayers of 2DFGs by the symmetry of the Hamiltonian, irrespective of parameter choices. The Hamiltonian can be easily generalized from the Supplementary Eq.~(S2), and we present the eigenvalue for some of the relevant bulk bands as,
\begin{eqnarray}
E^{\pm}({\bf k}) &=& \frac{k^2}{2m^*}\pm\frac{1}{3X({\bf k})}\left[3\left(3\alpha^2({\bf k})+2t_z^2+3D({\bf k})\right)X^2({\bf k})
+6t_z^2\left(t_z^2+6D({\bf k})\right)X({\bf k}) \right. \nonumber\\
&&~~~~~~~~~~~~~~~~~~~~~~~~~~~~\left. + 6\left(3Y({\bf k})-2\right)t_z^6+54t_z^4D^2({\bf k})\right]^{1/2},\hspace{120pt}{\rm (S4)} \nonumber
\label{biRashbaE3}
\end{eqnarray}
where
\begin{eqnarray}
X({\bf k}) &=& t_z\left[-8t_z^2+36t_z D({\bf k}) +12Y({\bf k})\right]^{1/3},\nonumber\\
{\rm{and}}~~~~~~~~~~~~~~~Y({\bf k}) &=&D({\bf k})\left[-96D^4({\bf k})-39t_z^2D^2({\bf k})-12t_z^4\right].
\hspace{140pt}{\rm (S5)} \nonumber
\label{biRashbaX}
\end{eqnarray}
The important message of the Supplementary Eq.~(S4) is that there is a prefactor $1/X\sim 1/t_z(...)$ which enables an inverted band gap opening for a negative value of $t_z$. The resulting parity inversion in the valence band endows a non-trivial topological phase transition at a large parameter space (see Fig.~3 of main text). A visual proof of the inverted band gap can be obtained from the associated inverted band curvature of `dent' shape in the valence bulk band, as illustrated in Supplementary Fig. S1c. According to Fu-Kane criterion,[7] for such case gapless Dirac states appear at the surface. However, for most values of the parameters shown in Supplementary Fig.~S2, a finite gap at the surface still persists due to the finite size-effect, and the surface gap closing requires a characteristic number of Rashba-bilayers.
\newline\\
\noindent
$\bullet$~{\bf Dependence of bulk and surface band structure on various parameters}\\

For the benefit of understanding the role of each parameter on the bulk and surface band structures, and also for the material growth purpose, we present results for a systematic tuning of various parameters as a function of number of layers in Supplementary Fig.~S2. The same results present in Fig.~2 of the main text is reproduced here in Supplementary Fig. S2a and the parameters are listed in the left side of this panel. For each row in the figures below, a single parameter is tuned while the others are kept to the value listed in Supplementary Fig.~S2a.
\newline\\
\noindent
\blue{$\bullet$ $m^*$:} The effective mass $m^*$ of the electrons have any significant effect neither on the bulk and the surface band structure, nor on the slope of the emergent surface state. Since this quadratic term goes to zero at the $\Gamma$-term, it begins to play a role at higher energy, and with increasing $m^*$, it only slightly reduces the indirect bulk gap.
\newline\\
\noindent
\blue{$\bullet$ $\alpha_R$:} An added benefit of the present proposal is that the Rashba SOC does not significantly affect the bulk or the surface gaps (very minor contribution to the indirect bulk gap), in contrary to the 3D bulk materials in which the bulk insulator gap is mainly controlled by the SOC. However, as deduced in the main text in Eq.~3, $\alpha_{\rm R}$ determines the velocity of the Dirac fermions on the surface. Therefore, for even a smaller value of $\alpha_{\rm R}$, the `non-trivial' topological insulator and the Dirac surface state occurs, only the velocity and the linear dispersion energy span deceases with $\alpha_{\rm R}$. As tabulated in Ref.~28, in 1 or 1/3 monolayer of Bi, which is measured to give a 2DFG (not a `surface' state of a 3D bulk system), one can achieve $\alpha_{\rm R}$ to be as large as 2.5-3 eV\AA. Assuming that in our proposed Rashba bilayer, this strength gets reduced by 1/3 or 1/2, one can still achieve our parameter set.

An interesting case of inhomogeneous Rashba-spin-orbit coupling can be mentioned here. Local fluctuations of Rashba SOC can be both a blessing and a curse to our proposal. If the Rashba SOC is spatially modulated in such a way that both counter-helical Rashba states can be achieved within a single layer, this can replace the effort to create our proposed Rashba bilayers. In one-dimensional systems, the possibility of a negative/positive switching effect of spin polarization via tuning spatially modulated Rashba SOC is studied.[42]

On the other hands, if the local fluctuation of Rashba SOC is highly inhomogeneous, it can lead to a reduction of the spin-polarization and an uneven gap structure in each layer, which is a curse to the topological properties. However, if this gap opening can be reduced to an extent that topological properties are retained, inhomogeneous Rashba SOC can be useful. There are theoretical proposals that inhomogeneous Rashba SOC can induce non-trivial phenomena such as magnetoelectric effect, noncentrosymmetric superconductivity[43], which can be realized within the topological matrix.
\newline\\
\noindent
\blue{$\bullet$ $t_z$:} As we have shown in Fig. 3, the inter-bilayer overlap matrix element $t_z$ mainly controls the bulk insulating gap. Even for $t_z$ as small as 100 meV, we get an insulating gap of about 200 meV. It should be noted that the actual hopping energy can be improved in two ways. While $t_z$ is the overlap matrix-element, the actual electronic hopping is $E_z=t_z\exp{(ik_zd)}$, where $d$ is the interlayer distance. Therefore, by reducing $d$, one can enhance $E_z$ up to a critical limit for a given setup, which will not kill the opposite Rashba couplings between them. The second and more important mechanism is to increase the effective gap for a given value of parameters by increasing the number of bilayers, $N$, in the heterostructure.
\newline\\
\noindent
\blue{$\bullet$ $D_0$:} Unlike $t_z$, $D_0$ does not have any significant effect on the bulk gap, but it controls the surface gap. However, as shown in Fig. 3, surface gap can easily be closed by increasing the number of bilayers, $N$.  In our results, we have used the values of $t_z$ and $D_0$ as in the range of 10-300meV. This small value of hopping is easily achievable even for the case of a weak Van-der Waals interaction between the two layers.
\newline\\
\noindent
\blue{$\bullet$ $M$:} The Newton mass $M$ is associated with a quadratic momentum dependence, and thus does not participate in the gap opening, and it contributes at higher energy. Higher the value of $M$, better the band dispersion topology.

Finally, it should also be noted that the `not-trivial' bulk insulator and surface Dirac fermion properties are the manifestation of the symmetry invariance of the Hamiltonian, not a numerical result for a set of parameters. The results presented are calculated using a realistic parameter range, which are indeed achievable in realistic materials.

\clearpage
\newpage
\vskip0.25cm\noindent
{\Large{\bf Suplementary References}}\\

[41] Bychkov,  Y. A. \& Rashba, E. I. Properties of a 2D electron gas with lifted spectral degeneracy. {\em JETP Lett.} {\bf 39}, 78-81 (1984). 

[42] Gong, S. J., \& Yang, Z. Q. Spin filtering implemented through Rashba spin-orbit coupling and weak magnetic modulations. {\em J. App. Phys.} {\bf 102}, 033706 (2007).

[43] Aoyama, K. \& Sigrist, M. Model for magnetic flux patterns induced by the influence of in-plane magnetic fields on spatially inhomogeneous superconducting interfaces of LaAlO$_3$-SrTiO$_3$ bilayers. {\em Phys. Rev. Lett.} {\bf 109}, 237007 (2012).


\end{document}